\documentclass{article}
\usepackage{tabularx} 
\usepackage{booktabs}
\usepackage{PRIMEarxiv}
\usepackage{ragged2e} 
\usepackage[utf8]{inputenc} 
\usepackage[T1]{fontenc}    
\usepackage{hyperref}       
\usepackage{url}            
\usepackage{booktabs}       
\usepackage{amsfonts}       
\usepackage{nicefrac}       
\usepackage{microtype}      
\usepackage{lipsum}
\usepackage{fancyhdr}       
\usepackage{graphicx}       
\graphicspath{{media/}}     
\usepackage{booktabs} 
\usepackage{float}
\usepackage{amsmath}
\usepackage{subcaption}
\usepackage{placeins}

\usepackage{caption}

\title{MOZART: Ensembling approach for COVID-19 detection using chest X-Ray imagery
}

\author{
  Mohammed Shabo\\
  University of Khartoum\\
  Khartoum, Sudan\\
  \texttt{mohammedshabo98@gmail.com} \\
  \And
  Nazar Siddig\\
  University of Khartoum\\
  Khartoum, Sudan\\
  \texttt{nazarsiddig@gmail.com}
}

\begin{document}
\maketitle

\begin{abstract}

COVID-19, caused by the SARS-CoV-2 virus, has resulted in a global pandemic, significantly straining healthcare systems worldwide. Early and accurate detection is crucial for suppressing the spread of the virus and ensuring timely treatment. While reverse transcription polymerase chain reaction (RT-PCR) tests remain the gold standard for diagnosing COVID-19, their limited availability, long processing times, and extremely high false negative rate \cite{pecoraro2022estimate}, have prompted the exploration of alternative methods. In this context, chest X-ray imaging has surfaced as a valuable diagnostic tool, offering a rapid, non-invasive approach to identifying COVID-19-related lung abnormalities.

 Although traditional convolutional neural networks (CNNs) can achieve impressive accuracy, there is still a pressing need for more robust and reliable solutions that minimize false positives and false negatives, especially in critical medical applications. To address this problem, we introduced the MOZART framework (Mohamed and Nazar Technique), an ensemble learning approach aimed at enhancing the virus detection. Within this framework, we trained three different CNN architectures—InceptionV3, Xception, and ResNet50—each on a balanced chest X-ray dataset containing 3,616 images of COVID-19 cases and 3,616 healthy images. Each model went through a separate pre-processing pipeline, such as normalizing inputs to a (-1,1) range. The dataset was divided into 70\% for training, 20\% for validation, and 10\% for testing. After the individual models finished generating their predictions, we concatenated the outputs and fed them into a shallow neural network. This network comes in two versions, MOZART1 and MOZART2, which reflect different experimental learning rates that outputs the final predictions.

Our results demonstrate that the MOZART ensemble framework outperforms individual CNN models in key metrics. The framework, including its sub-experiments (MOZART1 \& MOZART2) achieved the highest accuracy of 99.17\%, and the highest F1 score of 99.16\% with overall high precision and recall. The experiments also revealed that MOZART1 is optimal in contexts when minimizing false positives is prioritized, while MOZART2 is better suited for reducing false negatives. This work suggests that the MOZART framework is a valuable tool for improving reliability in AI-driven medical imaging tasks. Future work should explore its indulgence with other lung diseases to broaden its sample space.
\end{abstract}

\keywords{COVID-19 Detection \and Chest X-ray Imaging \and Deep Learning \and Convolutional Neural Networks \and Medical Imaging \and Image Classification}

\section{Introduction}
The COVID-19 outbreak has evolved to become one of the most severe public health crises in  recent years. The virus spreads like wildfire, the  reproduction number of COVID-19 went high up to 6.33 in places like Germany, indicating that each infected person infects six or more people on average \cite{linka2020reproduction}. As a result, the number of COVID-19 infections increased from a few hundred cases (mostly in China) in January 2020 to more than 770.3 million cases worldwide as of October 2024 \cite{owid2024,who2024}.  \\
Early diagnosis is crucial in the above-mentioned scenario to ensure patients receive appropriate  treatment while decreasing the burden on the healthcare system. COVID-19 remains a fatal disease due to a lack of early detection procedures around the world, as well as the presence of medical  preconditions such as cancer, chronic liver, lung, and renal disease, and diabetes. Despite the fact  that RT-PCR diagnosis procedures are widely available in most regions of the world, developing  countries cannot afford to test all of their citizens in a timely manner \cite{webmd2024}.

Millions of people died  as a result of this sickness in the years 2020–2023. COVID-19 vaccines are now developed in a  number of countries. The disease's timeliness has been significantly lowered as a result of these licensed immunizations. One of the major applications of Deep learning in radiology practices was the rapid detection of tissue skeletal abnormalities and the classification of diseases. The convolutional neural network has  proven to be one of the most important DL algorithms and the most effective technique in detecting  abnormalities and pathologies in chest radiographs \cite{rajpurkar2017chexnet}.

\section{Methodology}
The experiments of this study was conducted with well-known software and hardware components as can be shown in Table(\ref{tab:software and hardware}).
\label{sec:methodology}

\begin{table}[h]

\setlength{\tabcolsep}{30pt} 
\captionsetup{skip=10pt}
\centering

\begin{tabular}{cc}
    \toprule
    Software/Hardware & Details \\
    \midrule
    Python & version 3.9 \\
    Tensorflow & 2.16 \\
    Apple M1 chip & 7 GPU cores \\
    \bottomrule
\end{tabular}
   \caption{Software \& Hardware Utilized}
  \label{tab:software and hardware}

\end{table}

\subsection{Dataset}
\begin{figure}[h]  
    \centering
    \includegraphics[width=0.35\linewidth]{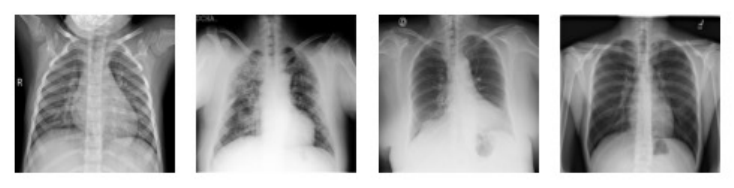}
    \caption{Chest XRay Dataset}
    \label{fig:dataset}
\end{figure}

\justifying
This experiment utilized a chest X-ray dataset shown in Figure(\ref{fig:dataset}) gathered by research teams from Qatar and Bangladesh \cite{kaggle2024}. The dataset includes 3,616 images of COVID-19 cases, along with 10,192 healthy, 6,012 lung opacity, and 1,345 viral pneumonia images. However, for this study, only the 3,616 COVID-19 images and 3,616 healthy images were used to ensure equal distribution and avoid bias toward either class. The selection process was randomized using Python's Random library. The final dataset, with images sized to 299x299 pixels, was split into training, validation, and testing sets in a 70\%, 20\%, and 10\% ratio, respectively.

\subsection{Proposed Approach}
\begin{figure}[H]
    \centering
    \includegraphics[width=0.35\linewidth]{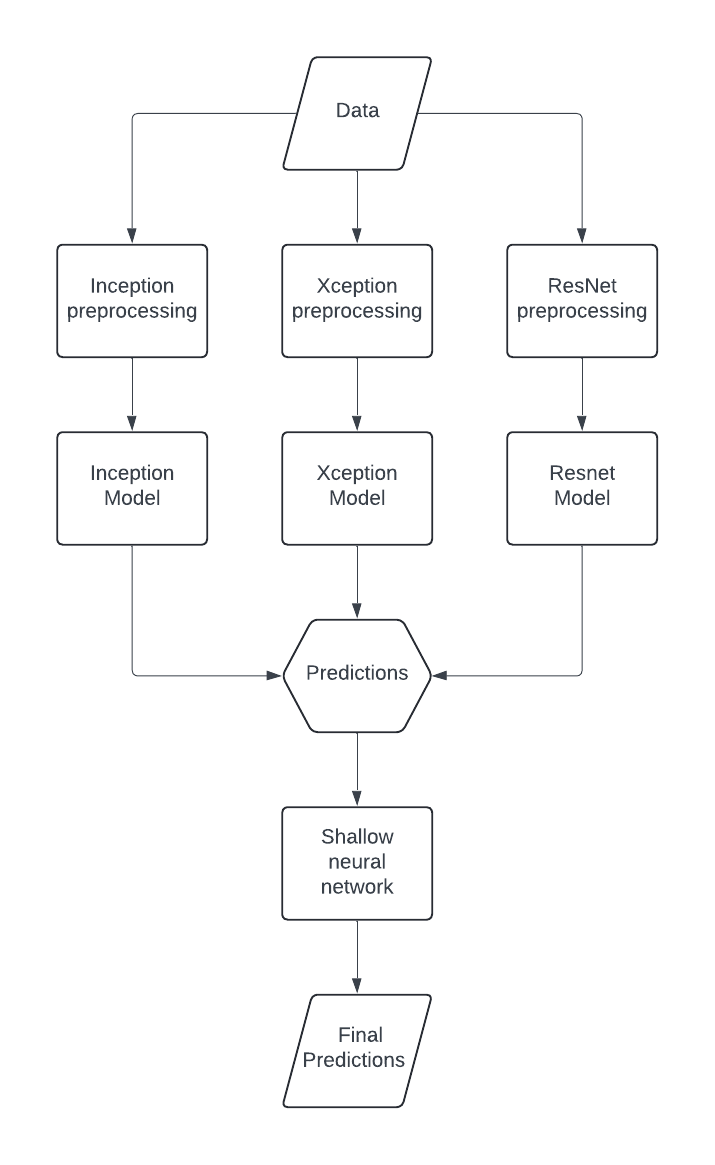}
    \caption{Proposed approach}
    \label{fig:approach}
\end{figure}

\justifying
In real-world scenarios, it's often preferable to base decisions on the opinions of multiple  expert individuals compromising the concept of "wisdom of the crowd"\cite{stehouwer2023collective}. The combined expertise of specialists contributes to a more reliable and well-rounded diagnosis. Following an analogous approach, our proposed methodshown in Figure(\ref{fig:approach}), dubbed MOZART (\textbf{Mo}hamed and Na\textbf{zar} \textbf{T}echnique), employs multiple benchmark CNN models. Each model is trained independently to make its own predictions, then these predictions are  concatenated and fed into a shallow neural network, enhancing the robustness of the overall architecture.

\subsection{Deep Learning Models}
\subsubsection{Inception\_v3} 
The Inception model is a strong model that can extract features with high accuracy and categorize  images using those features. The 48-layer deep architecture of is made up of 11  inception modules. Convolution filters, pooling layers, and the ReLU (Rectified Linear Unit)  activation function are all included in each inception module.\cite{szegedy2016rethinking}

\subsubsection{Resnet50}
Before ResNets, deep networks were built by stacking more layers, but this led to diminishing performance and high computational cost due to the vanishing gradient problem. As models became deeper, gradients would shrink, preventing effective training of the top layers. Previous solutions failed to fully resolve this issue. ResNets addressed it by introducing shortcut connections, which allow information to bypass layers, reducing training error in deeper networks. This idea is similar to Highway Networks, where layers can be skipped to improve information flow.\cite{he2016deep}

\subsubsection{Xception}
The Xception model is an evolution of the Inception architecture, specifically improving upon Inception V3. It replaces the traditional Inception modules with depthwise separable convolutions, allowing spatial and channel-wise filtering to be performed separately. This architectural shift results in a more efficient model that reduces the number of parameters while maintaining strong performance in feature extraction.\cite{chollet2017xception}

\subsection{Modifications on the deep learning Models}
The last layer was removed, and the following layers were added for regularization and to increase  the models performance: 

\begin{itemize}
    \item A Dense layer with 1024 Neurons
    \item Batch-normalization layer.
    \item  Dropout of (0.4)
    \item  A final layer with 1 neuron with a sigmoid activation function. 
\end{itemize}

\subsection{Shallow Neural Network}
\textbf{Dataset Subdivision}

The validation set was later further subdivided into a ratio of 80\% and 20\%  for the purpose of training and validating the last neural network in the architecture, with the testing set staying the same for the final evaluation of the model.

\textbf{Architecture}
\begin{itemize}
    \item Dense layer with 64 neurons with a ReLU activation function.
    \item Dropout layer with a value of (0.1)
    \item Dense layer with 32 neurons with a ReLU activation function.
    \item Dropout layer with a value of (0.1)
    \item One neuron layer with a Sigmoid activation function
\end{itemize}

\subsection{Data Augmentation}
Data augmentation was introduced as a method to increase the qualitative aspect of the samples in  our dataset \cite{perez2017effectiveness}, as well as to improve the reliability of the classifier and other performance metrics.

\begin{table}[h]

\setlength{\tabcolsep}{30pt} 
\captionsetup{skip=10pt}
\centering

\begin{tabular}{cc}
    \toprule
    Type & Details \\
    \midrule
    Shearing & 0.1 \\
    Zoom & 10\% \\
    rotation & 10\% \\
    Horizontal Flipping & True \\
    \bottomrule
\end{tabular}
   \caption{Data Augmentation}
  \label{tab:data augmentation}

\end{table}

The augmentation was conducted utilizing the (Image Data Generator) in Keras.

\subsection{Data Pre-processing}

The pre-processing techniques used in this experiment mirrored those applied when training models on it original ImageNet dataset, as these settings are more suited and used as a standard for the models. Each model underwent pre-processing steps tailored to its architecture. Table(\ref{tab:data preprocessing}) demonstrates the types of pre-processing.

\begin{table}[h]

\setlength{\tabcolsep}{30pt} 
\captionsetup{skip=10pt}
\centering

\begin{tabular}{cc}

    \toprule
    Model & Pre-processing \\
    \midrule
    Inception\_V3 & normalize input to [1,-1]  \\
    Xception & normalize input to [1,-1] \\
    ResNet50 & subtracting from the mean of the ImageNet dataset \\
    \bottomrule
\end{tabular}
   \caption{Data Pre-processing}
  \label{tab:data preprocessing}
  
\end{table}

\subsection{Optimization Algorithm}
The optimizer used throughout the experiments is the \textbf{Adam} Optimizer and the following parameters was found to be the most performant: \( \beta_1 = 0.9 \), \( \beta_2 = 0.999 \), \( \epsilon = 10^{-8} \), and \(\alpha = 0.0001\).
\subsection{Performance metrics}
\subsubsection{Accuracy}
It’s generally defined as the correctly classified samples to the whole samples of the dataset.
In this experiment it’s defined as the correctly classified positive and negative (normal and
COVID-19) samples to the \textbf{whole} dataset.
\[
\text{Accuracy} = \frac{TP + TN}{TP + TN + FP + FN}
\]
\subsubsection{Precision}
Precision is a measure of accuracy in the event that the testing result is positive. In other
words, how reliable it is if you test positive. True Positive/Total Classified Positive is used to
calculate precision.
\[
\text{Precision} = \frac{TP}{TP + FP}
\]
\subsubsection{Recall}
This metric counts how many positive cases we identified out of the total number
of positive cases. recall is calculated as True Positive/Total Actual Positive.
\[
\text{Recall} = \frac{TP}{TP + FN}
\]
\subsubsection{F1 Score}
F1 is a precision and recall function. F1 Score is required when attempting to find a balance
between Precision and Recall.
\[
F1 = 2 \times \frac{\text{Precision} \times \text{Recall}}{\text{Precision} + \text{Recall}}
\]

\section{Results and Discussions}
The binary classification (Normal vs. COVID-19) was investigated based on X-Ray imagery using transfer learning of the deep learning architectures, in order to determine the top performing architectures to move further with our proposed approach. To begin, the performance of  each deep learning architecture was assessed on its own using the performance metrics, then these architectures were used by concatenating their results and feeding it as an input to the final neural network. 
The models conducted throughout the experiment were the best architectures among many, and The callback  function provided by Keras was utilized to monitor the model at each epoch and save the weights of the best performing weights through the entire training process to be used later based on a pre-specified metric  which was chosen to be the (validation loss) as through experiments we concluded that it  correlates well to the other performance metrics. also we used a technique that reduces the learning rate when the monitored validation loss is plateaued and refuses to improve for a period of 5 epochs.

\subsection{Inception\_V3}
In this model, no resizing was done as this model takes images of input shape (299*299) which is the original resolution of the image so no re-sizing is needed. to make use of pre-trained parameters, only the layers from (Mixed7) and onwards were trained.

\begin{figure}[H]
    \centering
    \begin{subfigure}[b]{0.45\textwidth}
        \centering
        \includegraphics[width=\linewidth]{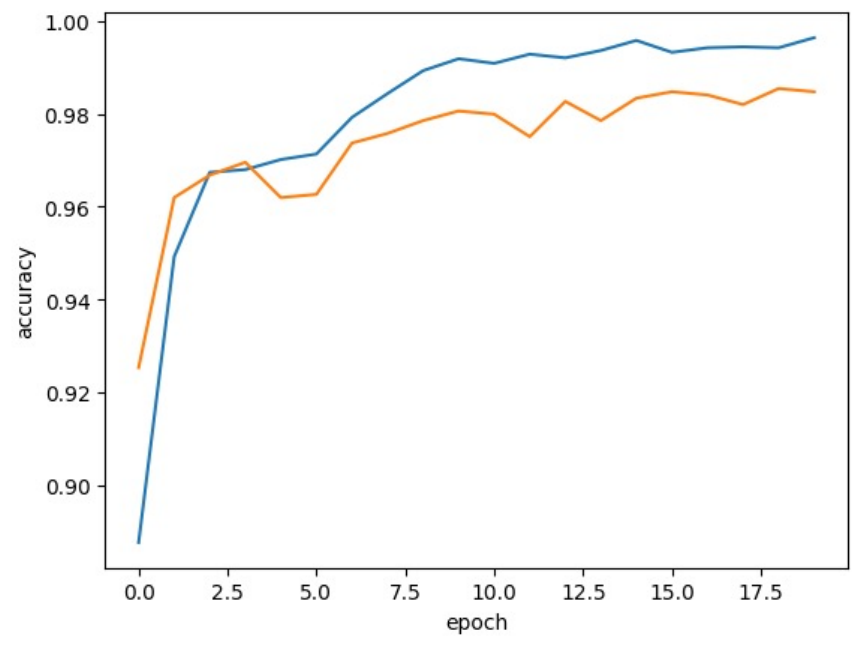}
        \caption{Performance plot}
        \label{fig:inception Performance plot}
    \end{subfigure}
    \hfill
    \begin{subfigure}[b]{0.45\textwidth}
        \centering
        \includegraphics[width=\linewidth]{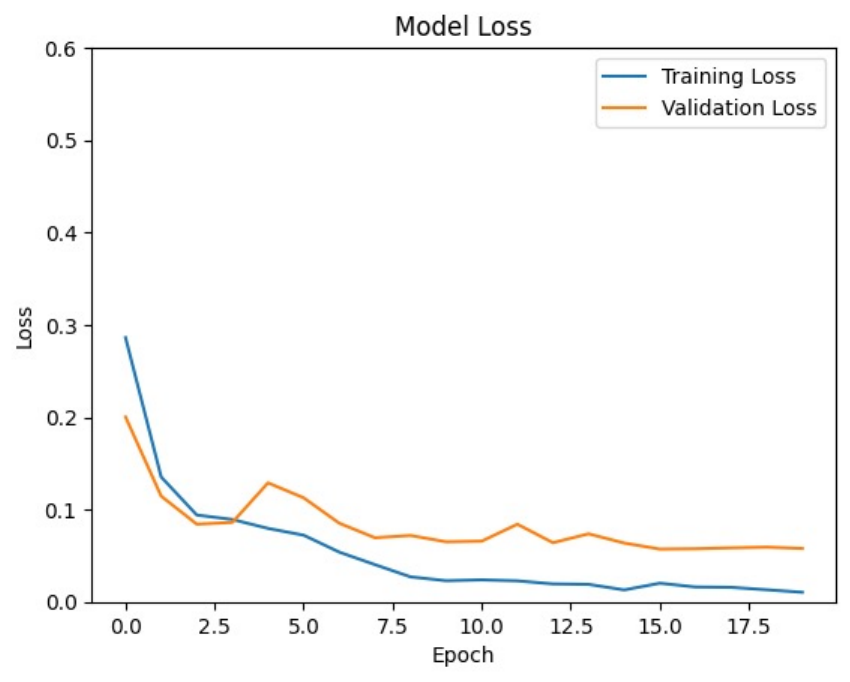}
        \caption{Learning Curve}
        \label{fig:inception learning curve}
    \end{subfigure}
    \caption{ Performance and Learning Curve}
    \label{fig:Inception side by side plots}
\end{figure}

\subsubsection{Training Log Analysis}

The training log of the Inception V3 model indicates a steady improvement in performance over the epochs. Below is a brief analysis of the log:

\textbf{Initial Epochs}
The model starts with a relatively lower accuracy of around 83\% in the first epoch and improves significantly to about 94\% by the second epoch.

\textbf{Learning Rate}
The learning rate is initially set to \(1.0 \times 10^{-4}\) for the first few epochs and is later reduced to \(2.0 \times 10^{-5}\). This adjustment helped stabilize learning and allowed stable steps towards convergence.

\textbf{Performance Metrics}
By the 20th epoch, the training accuracy reached approximately 99.62\%, while the validation accuracy standed at around 98.48\%. with the callback function monitoring the training returning the best performing weight according to the \textbf{validation loss}, the final returned weights were recorded on the 16th epoch achieving the following performance:

\begin{table}[h]

\setlength{\tabcolsep}{30pt} 
\captionsetup{skip=10pt}
\centering

\begin{tabular}{cccc}
    \toprule
    Metric & Training & Validation & Testing \\
    \midrule
    Accuracy & 99.45\% & 98.48\% & 98.48\% \\
    Precision & 99.36\% & 98.88\% & 96.95\% \\
    Recall & 99.55\% & 98.06\% & 99.78\% \\   
    
    \bottomrule
\end{tabular}
\caption{Performance Metrics for the Inception V3 Model}
\label{tab:inception_metrics}

\end{table}

\subsection{Xception Model}

\begin{figure}[H]
    \centering
    \begin{subfigure}[b]{0.45\textwidth}
        \centering
        \includegraphics[width=\linewidth]{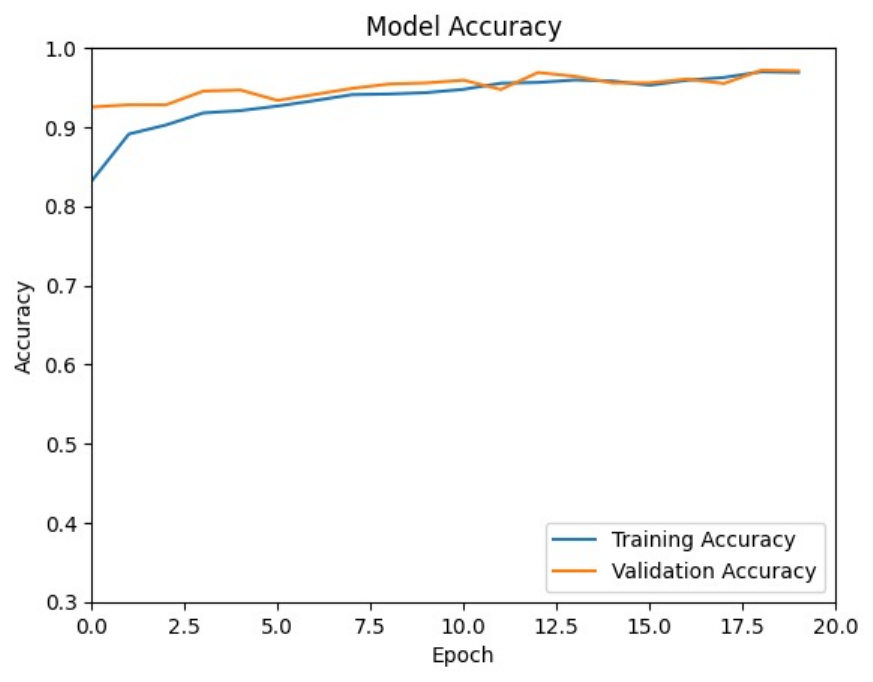}
        \caption{Performance plot}
        \label{fig:Xception Performance plot}
    \end{subfigure}
    \hfill
    \begin{subfigure}[b]{0.45\textwidth}
        \centering
        \includegraphics[width=\linewidth]{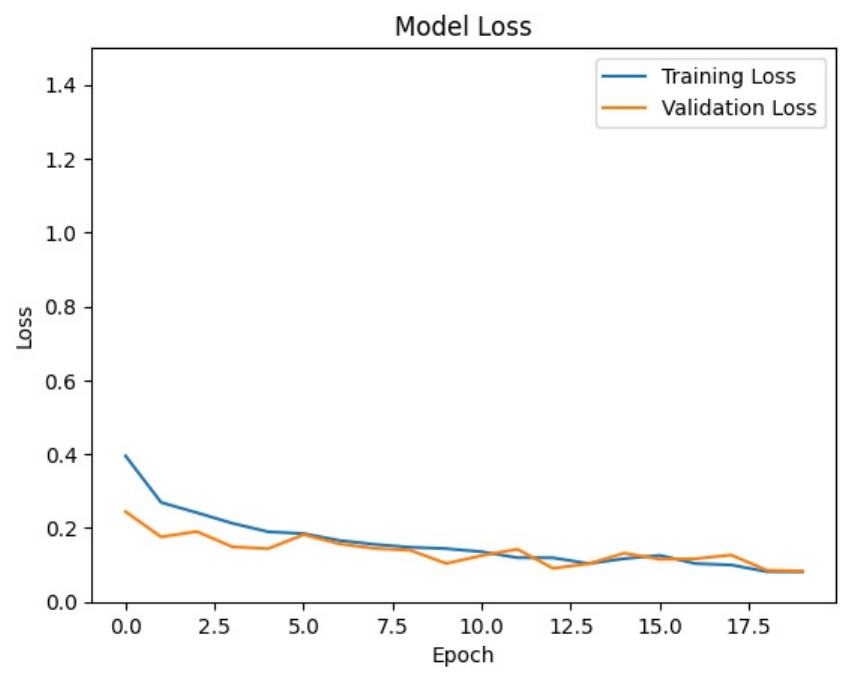}
        \caption{Learning Curve}
        \label{fig:Xception learning curve}
    \end{subfigure}
    \caption{Performance and Learning Curve}
    \label{fig:Xception side by side plots}
\end{figure}

Figure(\ref{fig:Xception side by side plots}) depicts The model's consistent improvement over epochs, with the performance metrics steadily increasing while the loss values steadily decrease, indicating good learning progression and convergence.

\subsubsection{Training Metrics}
\begin{itemize}
    \item Accuracy rises from 81.80\% to 97.03\%.
    \item Precision and Recall starts low but improve to 97.20\% and 96.93\% respectively by the final epoch.
    \item Loss steadily decreases, starting at 0.4711 and reaching 0.0789.
\end{itemize}

\subsubsection{Validation Metrics}
\begin{itemize}
    \item Accuracy rises to 97.10\% from an initial value of 92.53\%.
    \item Precision and Recall see consistent improvement, achieving values of 98.16\% and 95.99\%, respectively.
    \item Loss reduces significantly from 0.2444 in Epoch 1 to 0.0834 in Epoch 20.
\end{itemize}

\subsubsection{Learning Rate}
\begin{itemize}
    \item The learning rate starts at \(1.0 \times 10^{-4}\) for most epochs and reduces to \(2.0 \times 10^{-5}\) after Epoch 19.
\end{itemize}

\begin{table}[h]
\setlength{\tabcolsep}{20pt} 
\captionsetup{skip=10pt}
\centering
\begin{tabular}{lccc}
    \toprule
    \textbf{Metric} & \textbf{Training} & \textbf{Validation} & \textbf{Testing} \\
    \midrule
    Accuracy  & 97.03\% & 97.10\% & 95.29\% \\
    Precision & 97.20\% & 98.16\% & 91.96\% \\
    Recall    & 96.93\% & 95.99\% & 99.47\% \\

    \bottomrule
\end{tabular}
\caption{Performance Metrics of the Xception Model}
\label{tab:results_xception}
\end{table}

\subsection{ResNet}
The model shows a relatively steady improvement process and returning the best performing weights at epoch 20 with the weights achieving the performance in Table(\ref{tab:results_resnet50}).

\begin{figure}[H]
    \centering
    \begin{subfigure}[b]{0.45\textwidth}
        \centering
        \includegraphics[width=\linewidth]{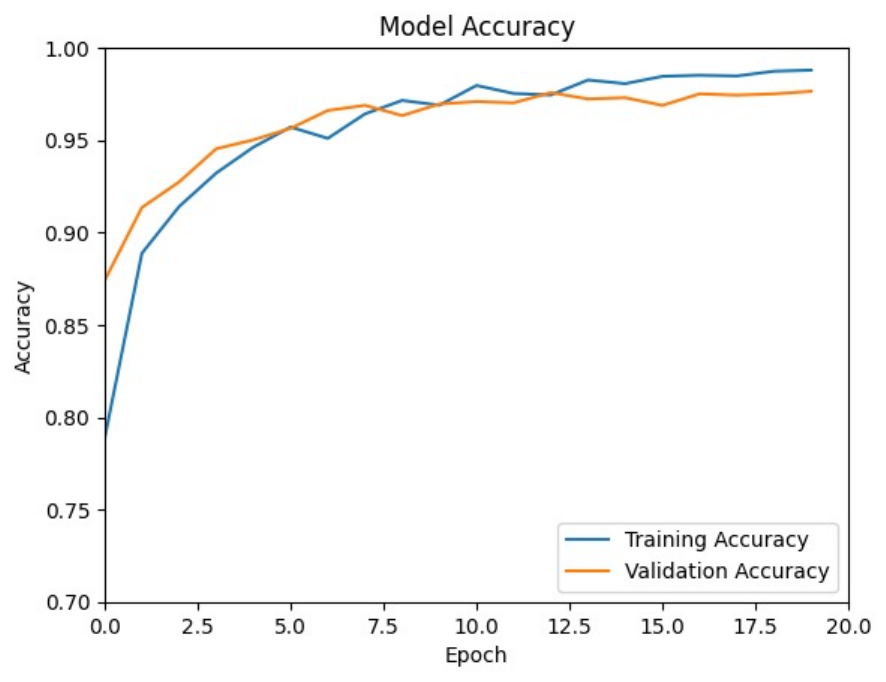}
        \caption{Performance plot}
        \label{fig:ResNet Performance plot}
    \end{subfigure}
    \hfill
    \begin{subfigure}[b]{0.45\textwidth}
        \centering
        \includegraphics[width=\linewidth]{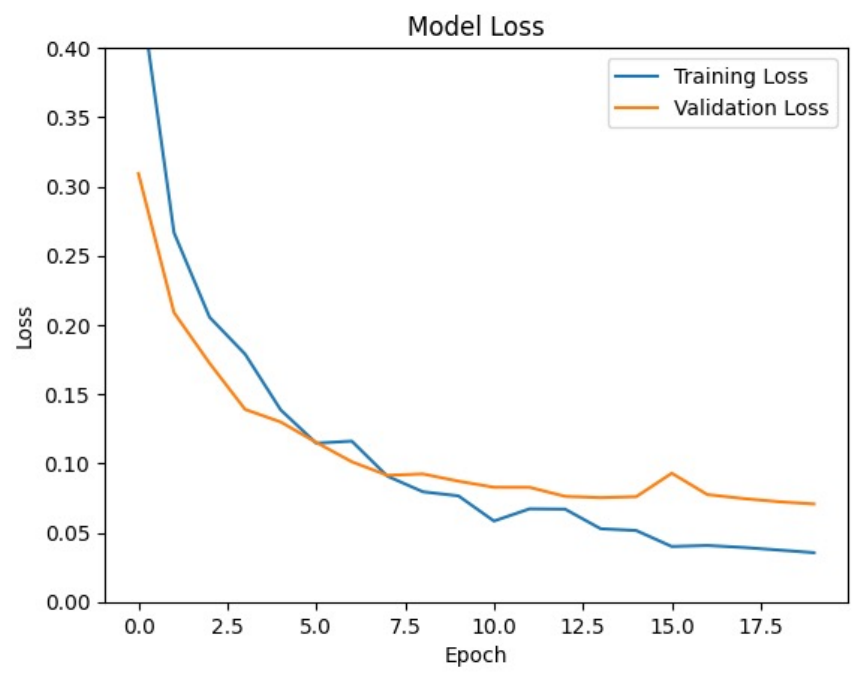}
        \caption{Learning Curve}
        \label{fig:ResNet learning curve}
    \end{subfigure}
    \caption{Performance and Learning Curve}
    \label{fig:Resnet side by side plots}
\end{figure}

\subsubsection{Training Metrics}
    \begin{itemize}
        \item Accuracy rises from 70.42\% in Epoch 1 to 98.91\% by the final epoch.
        \item Precision and Recall start lower at 71.57\% and 70.10\%, respectively, but improve to 98.94\% and 98.88\% by Epoch 20.
        \item Loss decreases significantly from 0.6073 in Epoch 1 to 0.0352  by the final epoch, indicating strong learning and convergence.
    \end{itemize}

\subsubsection{Validation Metrics}
    \begin{itemize}
        \item Accuracy improves from 87.41\% in Epoch 1 to 97.65\% by the final epoch.
        \item Precision and Recall also see improvement, starting at 92.60\% and 81.33\%, respectively, and reaching 97.91\% and 97.37\% by Epoch 20.
        \item Validation Loss decreases from 0.3093 to 0.0707 over the course of training, showing strong generalization and model fine-tuning.
    \end{itemize}

\subsubsection{Learning Rate}
- The learning rate started at 1.0e-04 and decreased until the best weights were captured with a learning rate of  \(5.0 \times 10^{-6}\)

\begin{table}[h]
    \setlength{\tabcolsep}{20pt} 
    \captionsetup{skip=10pt}
    \centering
    \caption{Training, Validation, and Testing Results of the ResNet50 Model}
    \begin{tabular}{lccc}
        \toprule
        \textbf{Metric}     & \textbf{Training (\%)} & \textbf{Validation (\%)} & \textbf{Testing (\%)} \\
        \midrule
        
        \textbf{Precision}  & 98.94\%  & 97.91\%  & 96.12\% \\
        \textbf{Recall}     & 98.88\%  & 97.37\%  & 99.49\% \\
        \textbf{Accuracy}   & 98.91\%  & 97.65\%  & 97.65\% \\
        \bottomrule
    \end{tabular}
    \label{tab:results_resnet50}
\end{table}

\subsection{MOZART}
Due to the relatively small number of parameters we were able to conduct two experiments on the shallow neural network with two different learning parameters.

\subsubsection{First Experiment}
\textbf{Learning rate of \(1 \times 10{-5}\)}

\begin{figure}[H]
    \centering
    \begin{subfigure}[b]{0.45\textwidth}
        \centering
        \includegraphics[width=\linewidth]{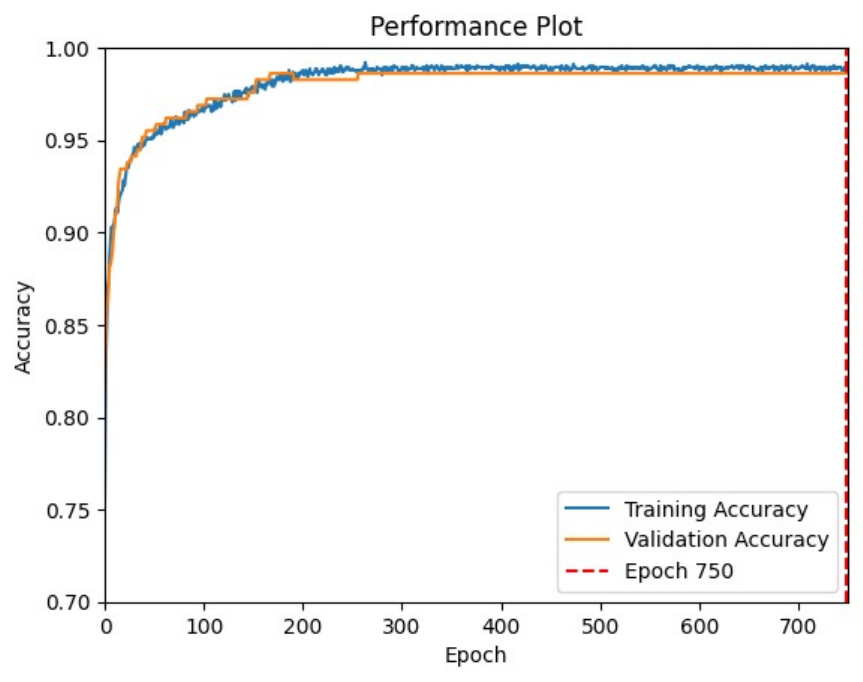}
        \caption{Performance plot}
        \label{fig:mozart Performance plot}
    \end{subfigure}
    \hfill
    \begin{subfigure}[b]{0.45\textwidth}
        \centering
        \includegraphics[width=\linewidth]{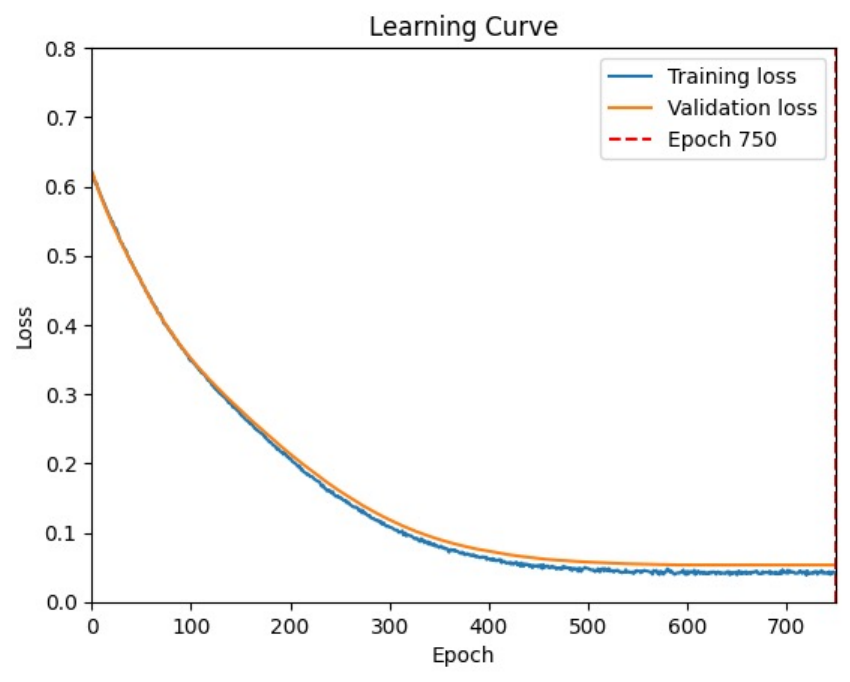}
        \caption{Learning Curve}
        \label{fig:mozart learning curve}
    \end{subfigure}
    \caption{Performance and Learning Curve}
    \label{fig:mozart side by side plots}
\end{figure}

\textbf{Training Analysis} \\
We can clearly observe in Figure(\ref{fig:mozart side by side plots}) the that the model is learning slowly because of the low learning rate, but the oscillations are quite modest when compared to the other models reflecting a more stable learning process. Also, we can observe that the decrease in loss after epoch 400 is rather relatively small when compared to preceding epochs. Furthermore, the gap between training and validation loss is relatively modest, implying minimal over-fitting as shown in Figure(\ref{fig:mozart learning curve}), achieving the results in Table (\ref{tab:metrics_1})

\begin{table}[h]
    \setlength{\tabcolsep}{20pt} 
    \captionsetup{skip=10pt}
    \centering
    \caption{First experiment metrics}
    \begin{tabular}{lccc}
        \toprule
        \textbf{Metric}     & \textbf{Training} & \textbf{Validation} & \textbf{Testing} \\
        \midrule
        Accuracy   & 98.91\%  & 98.62\%  & 99.17\% \\
       Precision  & 99.16\%  & 97.93\%  & 100.00\% \\
        Recall    & 98.71\%  & 99.30\%  & 98.34\% \\
        F1 Score   & 98.93 \%     & 98.61\%      & 99.16\% \\
        \bottomrule
    \end{tabular}
    \label{tab:metrics_1}
\end{table}

\subsubsection{Second Experiment}
\textbf{Learning rate of \(1 \times 10{-4}\)}

\begin{figure}[H]
    \centering
    \begin{subfigure}[b]{0.45\textwidth}
        \centering
        \includegraphics[width=\linewidth]{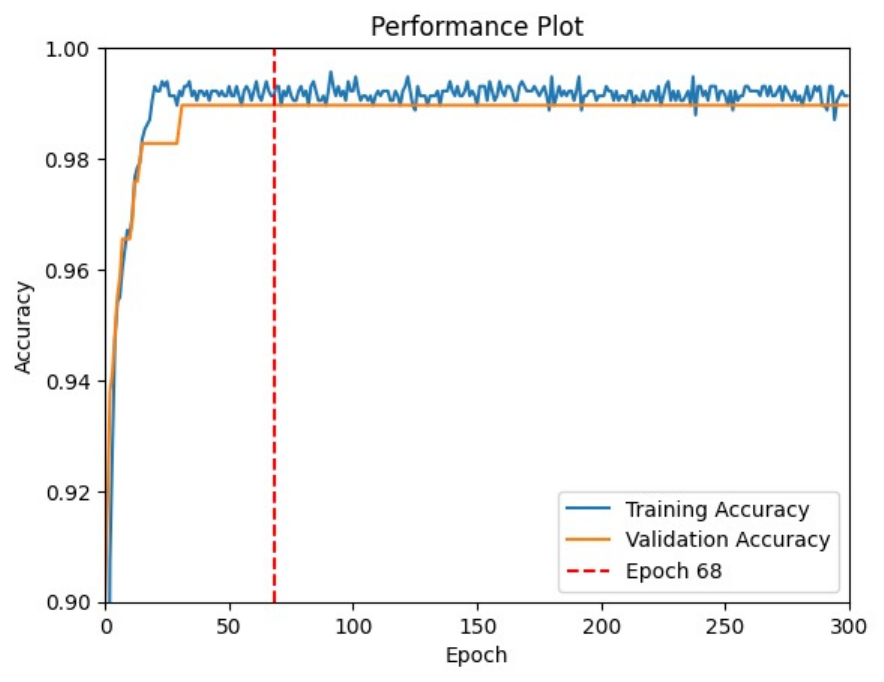}
        \caption{Performance plot}
        \label{fig:second mozart Performance plot}
    \end{subfigure}
    \hfill
    \begin{subfigure}[b]{0.45\textwidth}
        \centering
        \includegraphics[width=\linewidth]{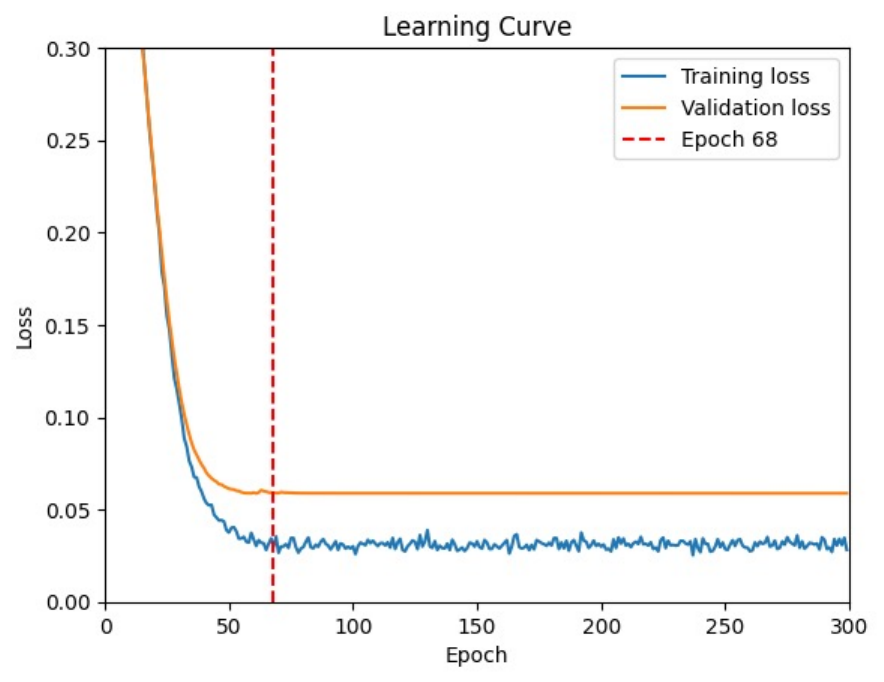}
        \caption{Learning Curve}
        \label{fig:second mozart learning curve}
    \end{subfigure}
    \caption{Performance and Learning Curve}
    \label{fig:second mozart side by side plots}
\end{figure}

\textbf{Training analysis}\\
in this experiment, by having 10 times more the learning rate we decreased the number of epochs to only 300 to counter for the fast convergence procedure as shown in Figure(\ref{fig:second mozart learning curve}) compared the previous experiment shown in Figure(\ref{fig:mozart learning curve}). with this setting we can observe that the rate of change in the loss becomes very subtle after the 50th epoch rather than the 400 epoch with a smaller learning rate in the previous experiment achieving also a very high metrics with low computing time as demonstrated below in Table (\ref{tab:metrics_2}):

\begin{table}[h]
     \setlength{\tabcolsep}{20pt} 
    \captionsetup{skip=10pt}
    \centering
    \caption{Second experiment metrics}
    \begin{tabular}{lccc}
        \toprule
        \textbf{Metric}     & \textbf{Training} & \textbf{Validation} & \textbf{Testing} \\
        \midrule
        Accuracy  & 99.35\%  & 98.97\%  & 99.17\% \\
        Precision  & 99.44\%  & 97.95\%  & 99.72\% \\
        Recall     & 99.23\%  & 100\%   & 98.61\% \\
        F1 Score   & 99.33\%      & 98.96\%      & 99.16\% \\
        \bottomrule
    \end{tabular}
    \label{tab:metrics_2}
\end{table}
\textbf{General Comparisons}
\newline
Table(\ref{tab:metrics_3}) demonstartes the performance metrics of all the experiments done in this study.
\begin{table}[h]
     \setlength{\tabcolsep}{10pt} 
    \captionsetup{skip=10pt}
    \centering
    \caption{Performance Comparison of individual Models and MOZART Framework}
\begin{tabular}{lccccc}
    \toprule
    \textbf{Metric}     & \textbf{Inception} & \textbf{Xception} & \textbf{ResNet} & \textbf{MOZART1} & \textbf{MOZART2}\\
    \midrule
    Accuracy  & 98.47\%  & 95.29\%  & 96.12\%  & 99.17\%  & 99.17\% \\
    Precision & 96.95\%  & 91.96\%  & 99.49\%  & 100\%    & 99.72\% \\
    Recall    & 99.78\%  & 99.47\%  & 97.65\%  & 98.34\%  & 98.61\% \\
    F1 Score  & 98.34\%  & 95.57\%  & 98.56\%  & 99.16\%  & 99.16\% \\
    \bottomrule
\end{tabular}

    \label{tab:metrics_3}
\end{table}

\section{Conclusion}
In this research, we introduced an ensemble technique for COVID-19 detection from chest X-ray images, using the MOZART (Mohamed and Nazar Technique) framework. This framework combines the strengths of multiple independent CNN models (Inception, Xception, and ResNet) into another neural network, which yielded significant improvements in diagnostic accuracy and other key metrics compared to the individual models.

When comparing the performance of the individual models, InceptionV3 demonstrated the highest accuracy at 98.47\%, with Xception and ResNet trailing at 95.29\% and 96.12\%, respectively. However,when further breaking down the accuracy metric, ResNet achieved the highest precision at 99.49\%, while Inception showed the highest recall at 99.78\%. Despite their individual merits, the ensemble approach proved to be more robust. The MOZARTx predictions, proved to have the best combination of high precision and high recall showing the highest f1 score among all the individual models, with both MOZART1 \& MOZART2 scoring 99.16\% on the f1 metric and the individual models scoring 98.34\%, 95.57\& , and 98.56\% for inception, Xception, and ResNet respectively. MOZART1 displayed the highest precision of 100\% but lagged a bit on the recall metric when compared with MOZART2, MOZART1 scoring 98.34\% and MOZART2 scoring 98.61\%.

Through two experiments on the shallow neural network using different learning rates, MOZART1 yields the same accuracy and F1 score as MOZART2, while it has a higher precision, making it more performant in contexts when minimizing false positives is prioritized. on the other hand, MOZART2 has a higher recall, indicating it is more efficient at reducing false negatives. Thus, each model can be utilized in different settings based on the specific needs for precision or recall.

The results of our experiments show that aggregating the predictions from multiple CNN models significantly enhances diagnostic performance. The MOZART ensemble approach can serve as a valuable tool in AI-driven medical imaging, providing more reliable COVID-19 detection. Future work may focus on introducing this architecture to include more lung diseases to increase the qualitative aspect and broaden the sample space. Additionally, exploring additional models to further validate the robustness and generalizability of the proposed technique.

\bibliographystyle{unsrt}  

\bibliography{references}

\end{document}